\documentclass{iaus}
\usepackage{graphicx}

\title[LSD and AMAZE]
{LSD and AMAZE: the mass-metallicity relation at z$>3$}

\author[F. Mannucci \& R. Maiolino]   
{F. Mannucci$^1$
\and R. Maiolino$^2$
\thanks{on behalf of the LSD and AMAZE collaborations},
}

\affiliation{
$^1$ INAF - IRA\\ 
	Largo E. Fermi 5, I-50125, Firenze, Italy \\
	email: {\tt filippo@arcetri.astro.it}\\
$^2$ INAF - OAR, Roma 
}

\pubyear{2008}
\volume{255}
\pagerange{1--5}
\jname{Low-Metallicity Star Formation: From the First Stars to Dwarf Galaxies}
\editors{L.K. Hunt, S. Madden \& R. Schneider, eds.}

\begin{document}

\maketitle

\begin{abstract}
We present the first results on galaxy metallicity evolution at z$>$3
from two projects, 
LSD (Lyman-break galaxies Stellar populations and Dynamics)
and AMAZE (Assessing the Mass Abundance redshift Evolution). 
These projects use deep near-infrared 
spectroscopic observations of a sample of $\sim$40 LBGs
to estimate the gas-phase metallicity from the emission lines. 
We derive the
mass-metallicity relation at z$>$3 and compare it with the same relation at
lower redshift. Strong evolution from z=0 and z=2 to z=3 is observed,
and this finding puts strong constrains  on the models of galaxy evolution.
These preliminary results show that the effective oxygen yields 
does not increase with stellar mass, implying that
the simple outflow model does not apply at z$>$3.

\end{abstract}

\firstsection 

\section{Introduction}

Metallicity is one the most important property of galaxies, and its study
is able to shed light on the detail properties of galaxy formation.
It is an integrated property, not related to the present level
of star formation the galaxy, but rather to the whole past history 
of the galaxy. In particular, metallicity is sensitive to the fraction of
baryonic mass already converted into stars, i.e., to the evolutive stage
of the galaxy.
Also, metallicity is effected by the presence of inflows and outflows, i.e.,
by feedback processes and the interplay between the forming galaxy
and the intergalactic medium.

It is well known that local galaxies follow a well-defined mass-metallicity 
relation, where galaxies with larger stellar mass have higher metallicities
(\cite{tremonti04,lee06}).
The origin of the relation is uncertain because several effects can be, 
and probably are, active. 
It is well known that in the local universe starburst galaxies
eject a significant fraction of metal-enriched gas into the intergalactic 
medium because of the energetic feedback from exploding SNe, both 
core-collapse and, possibly, type Ia (\cite{mannucci06}). Outflows are 
expected to be more important in low-mass galaxies, where the 
gravitational potential is lower and a smaller fraction of gas is retained.
As a consequence, higher mass galaxies are expected to be more metal rich
(see, for example, \cite{edmunds90,garnett02}). A second possibility is related
to the well know effect of ``downsizing'' (e.g., \cite{cowie96}), 
i.e., lower-mass
galaxies form their stars later and on longer time scales. At a given time, 
lower mass galaxies have formed a smaller fraction of their stars,
therefore are expected to show lower metallicities.
Other possibilities exist, for example some properties of star formation,
as the initial mass function (IMF), could change systematically 
with galaxy mass (\cite{koppen07}).

All these effects have a deep impact on galaxy formation, and the knowledge 
of their relative contributions is of crucial importance.
Different models have been built to reproduce the shape 
of the mass-metallicity relation in the local universe, and different 
assumptions produce divergent predictions
at high redshifts (z$>$2). To explore this issue several groups
have observed the mass-metallicity relation in the distant universe, 
around z=0.7 (\cite{savaglio05}) and z=2.2 (\cite{erb06}). They have found 
a clear evolution with cosmic time, 
with metallicity for a given stellar mass decreasing with
increasing redshift. 

For several reasons, it is very interesting to explore even higher redshifts.
The redshift range at z$\sim$3--4 is particularly interesting:  it is before 
the peak of the cosmic star formation density (see, for example, 
\cite{mannucci07}), only a small fraction 
($~$15\%, \cite{pozzetti07}) of the total stellar mass have already
been created, the number of mergers among the galaxies is much larger
than at later times (\cite{conselice07}). As a consequence, the prediction 
of the different models tend to diverge above z=3, and it is important to 
sample this redshift range observationally. The observations are really 
challenging because of the faintness of the targets and the precision 
required to obtain a reliable metallicity. Nevertheless, the new 
integral-field unit (IFU) instruments on 8-m class telescopes
are sensitive enough to allow for the project.

\section{LSD and AMAZE}

Metallicity at z$\sim$3 can be obtained by measuring the fluxes of the main
optical emission lines ([OII], H$\beta$, [OIII], H$\alpha$), whose 
ratios have been calibrated against metallicity
in the local universe (\cite{nagao06,kewley08}). Of course, this method 
can be applied only to line-emitting galaxies, i. e., 
to low-extinction, star-forming galaxies, whose line can be seen 
even at high-redshifts. In contrast, the gas metallicity of more quiescent 
and/or dust extincted galaxies, like EROs (\cite{mannucci02}), 
DRGs (\cite{franx03}), and SMGs (\cite{chapman05}), cannot be easily measured
at high redshifts
\footnote{Stellar metallicities, measured by absorption lines, can also 
be measured if enough observing time is provided, and will be subject 
of a future work.}.

For the {\bf AMAZE} project we observed
30 galaxies at z$\sim$3.3 from various sources 
(see \cite{maiolino08} for details).
We only selected galaxies having a good SPITZER/IRAC photometry 
(3.6--8 $\mu m$), an important piece of information to derive a reliable
stellar mass. These galaxies were observed, in seeing-limited mode,
with the integral-field unit (IFU)
spectrometer SINFONI on ESO/VLT, with integration times between 3 and 6 hours.
When computing line ratios, it is important that all the lines 
are extracted from exactly the same aperture, to avoid differential slit losses 
that could spoil the line ratio. 
In this, the use of an IFU is of great help.
We observed the H and K bands simultaneously with spectral resolution 
R$\sim$1500, providing 
a full coverage of all the most important lines. 
This paper is based on about 1/3 of the full data sample, while the remaining 
fraction is still under analysis.

For {\bf LSD}, we extracted 10 galaxies from the UV-selected 
Lyman Break Galaxies (LBG) sample by \cite{steidel04}. The aim of this 
project is not only to measure metallicities, but also to obtain 
spatially-resolved
spectra to measure dynamics and spectral gradients.
For this reason we used adaptive optics to obtain
diffraction-limited images and spectra in the  near-IR.
The target galaxies were chosen to be 
within 30'' of  bright foreground stars, needed to drive the
adaptive-optics system. 
As the presence of a nearby bright star is the only request, this
sample, albeit small, is expected to be representative of the full 
population of the LBGs. For LSD we used SINFONI with the same resolution 
of AMAZE and similar integration times.
About half of the LSD galaxies have been already analyzed.
The typical spatial resolutions is about 0.2". 

\begin{figure}[t]
\begin{center}
\includegraphics[width=11cm]{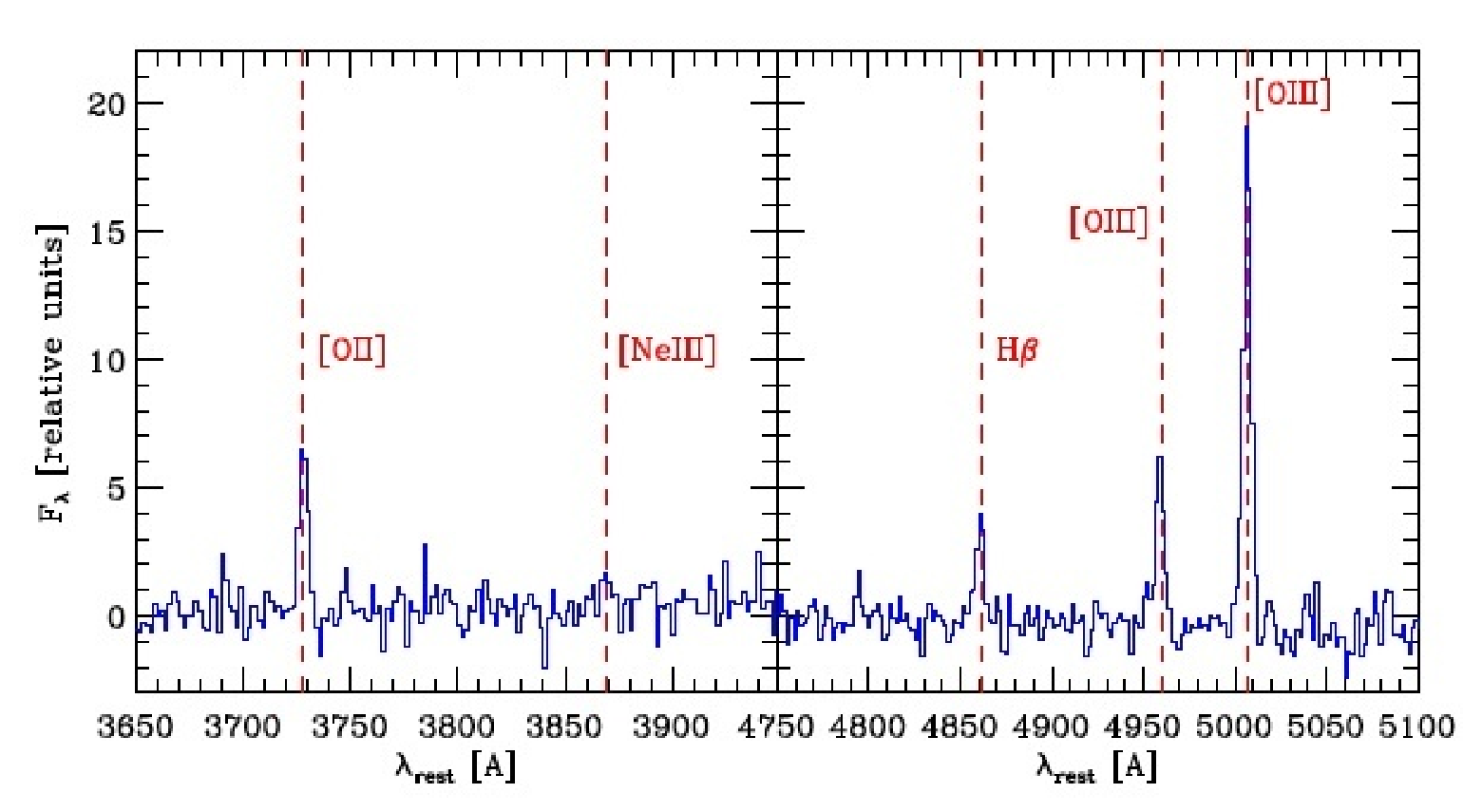} 

\includegraphics[width=8cm,angle=-90]{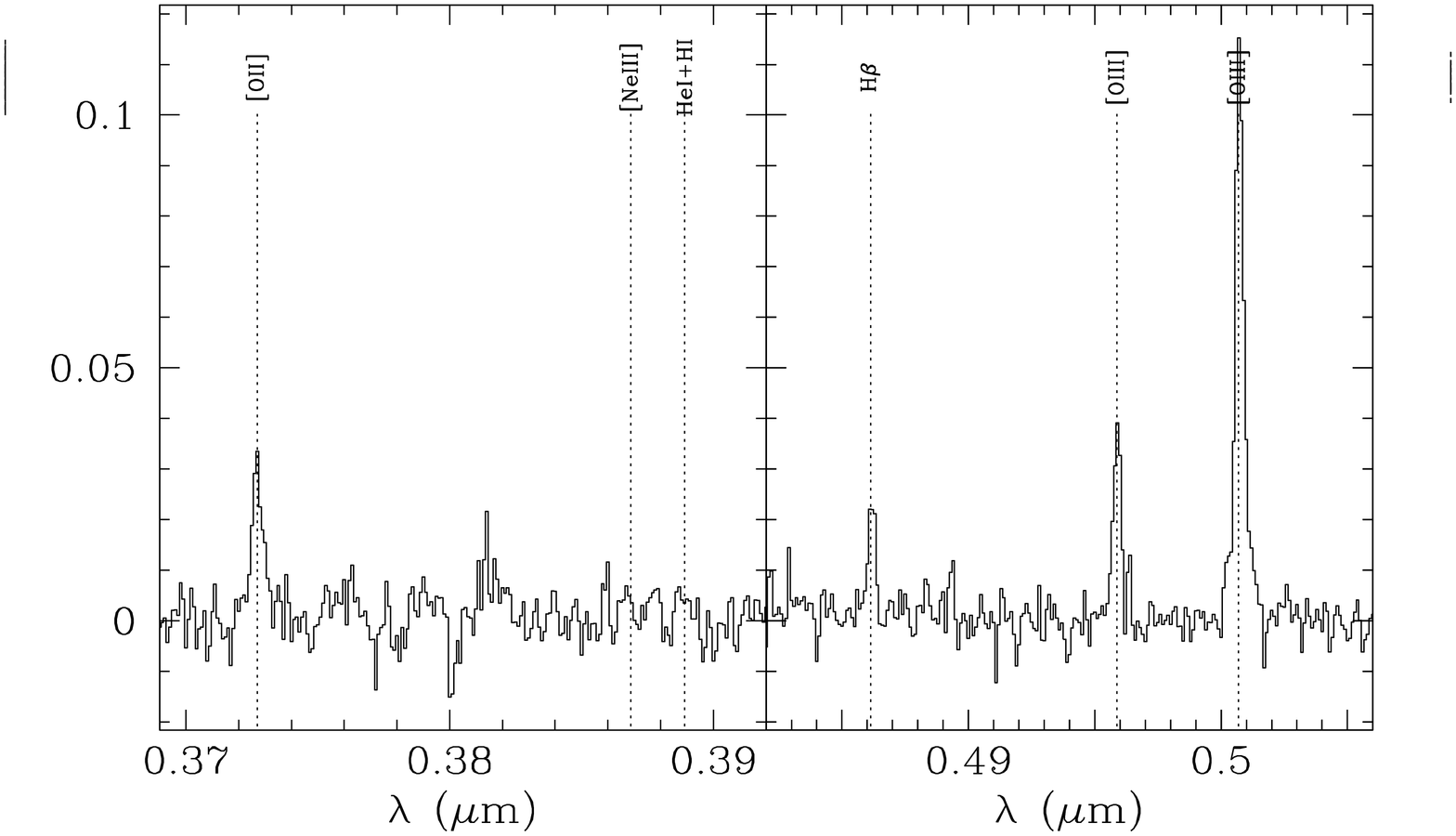} 
\caption{
Composite spectra from AMAZE (top) and LSD (bottom).
}
\label{fig1}
\end{center}
\end{figure}

Fig.\,\ref{fig1} shows the composite spectra of AMAZE (top) and LSD
(bottom).
For both AMAZE and LSD, the main optical lines are  detected
in most of the galaxies.  A few targets also show the [NeIII]3869 
line, and important piece of information to derive a robust measurement 
of metallicity (\cite{nagao06}).

\begin{figure}[b]
\begin{center}
\includegraphics[width=10cm]{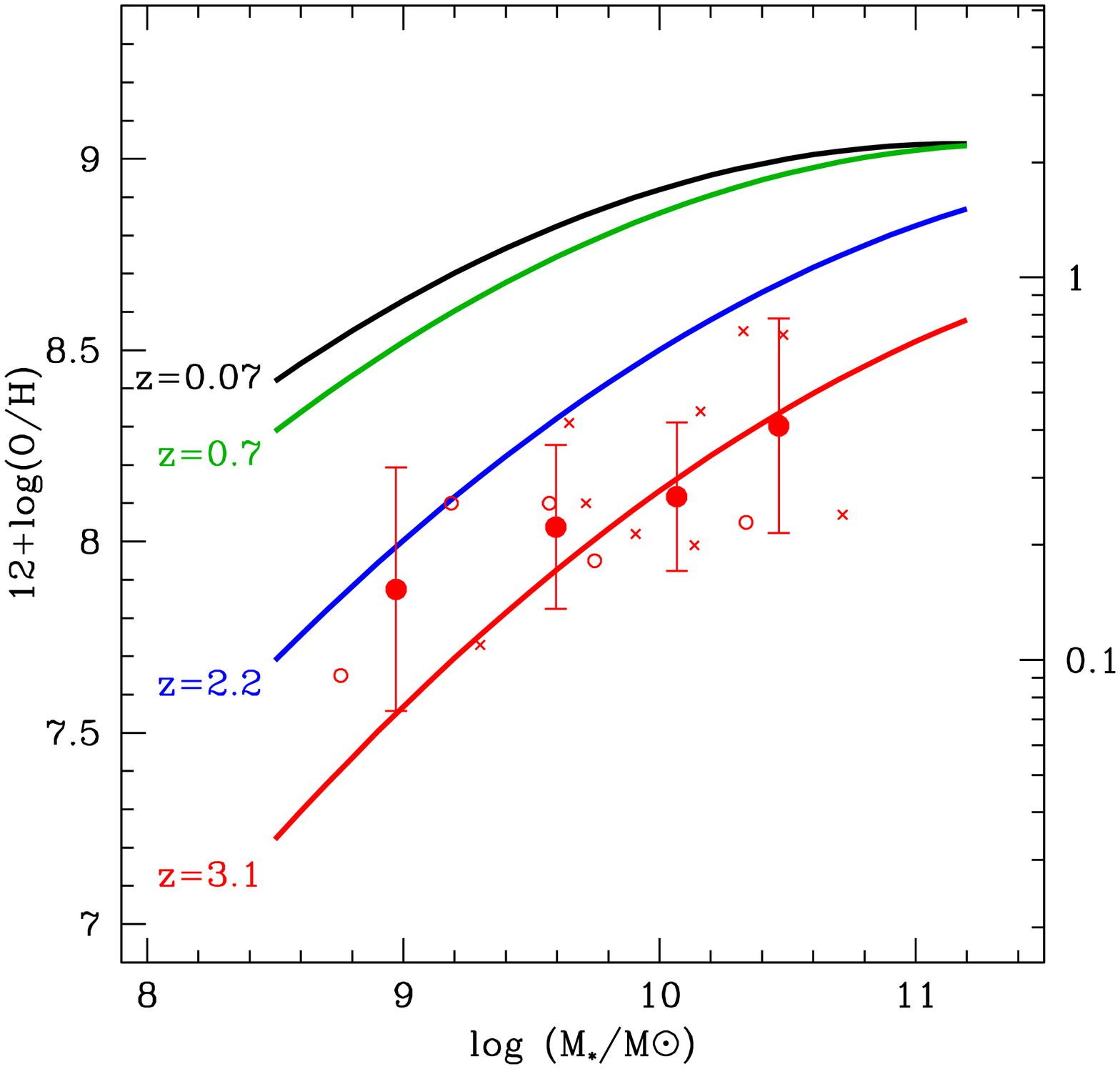} 
\caption{
Evolution of the mass-metallicity relation from z=0.07 (\cite{kewley08}),
to z=0.7 (\cite{savaglio05}), z=2.2 (\cite{erb06}) and z=3.1 (AMAZE+LSD).
All data have been calibrated to the same scale in order to make 
all the different results directly comparable.
Crosses show the AMAZE galaxies, empty dots the LSD galaxies. 
Solid dots with error bars show the average metallicity of the galaxies 
in each stellar-mass bin, with the associated dispersion.
The lines show quadratic fits to the data. The line corresponding to z=3.1
shows the fit in \cite[Maiolino et al. (2008)]{maiolino08}.
}
\label{fig2}
\end{center}
\end{figure}

\section{The Mass-metallicity relation and the effective yields}
 
Stellar masses are derived by fitting the spectral-energy distributions (SEDs)
with spectrophotometric models of galaxy evolutions, as detailed in
\cite{pozzetti07} and \cite{grazian07}. 
These fits also provide estimates of the age and dust
extinction of the dominant stellar population. The presence, for most of the
objects, of good IRAC photometry, allows the determination of reliable 
stellar masses as the SED is sampled up to the rest-frame J band. 

Metallicities are derived by a simultaneous fit of all the available line 
ratios, as explained  \cite{maiolino08}. 
In practice, the derived value is
determined by the R23 indicator or, similarly, 
by the [OIII]5007/H$\beta$ ratio, 
while the [OIII]5007/[OII]3727 ratio is used to discriminate between the two
possible branches of these ratios. The [NeIII]3869/[OII]3727 line ratio is
also very important, when both lines are detected. 
The uncertainties on metallicity
are due to both the spread of the calibration and on the observational error
on the line ratio, and on average amount to 0.2--0.3 dex.

Fig.\,\ref{fig2} shows the resulting
mass-metallicity relation, compared to the same relation as measured at lower
redshift.
A quantitative interpretation of this result will be given when the whole
data sample will be analyzed. Nevertheless, a strong metallicity
evolution can be seen, i.e., galaxies at z$\sim$3.1 have metallicities 
$\sim$6 times lower than galaxies of similar stellar mass in the local 
universe. 

The presence, at z$>$3, of galaxies with relatively high stellar masses 
(log(M/M$\odot$)=9-11) and low metallicity put strong constraints on the 
process dominating galaxy formation. 
Several published models (e.g., \cite{derossi07,kobayashi07}) 
cannot account for such a strong evolution, and the physical reason 
for this can be traced to be due to some inappropriate assumption,
for example about feedback processes or merging history. 
When taken at face value, 
some other models (e.g., \cite{brooks07,tornatore07}) provide a better 
match with the observations, but a meaningful comparison can only
be obtained by taking into account all the selection effects and
observational biases.\\

Important hints on the physical processes shaping the mass-metallicity
relation can be understood by 
considering the effective yields, i.e., the amount of metals produced 
and retained in the galaxy per unit mass of formed stars. 
If outflows are the main effect in shaping the mass-metallicity 
relation, then the effective yields are expected to increase with 
stellar mass because all the galaxies
have converted, on average, the same fraction of baryonic mass into stars,
but lower mass galaxies have lost a larger fraction of metals into the 
intergalactic medium. 
This is what is observed in the
local universe by \cite[Tremonti et al. (2004)]{tremonti04} 
(but not by \cite{lee06}).
For LSD and AMAZE, preliminary results show that the effective yields 
tend to decrease, rather than increase, with stellar mass,
as in \cite[Erb et al. (2006)]{erb06}. 
Such a decrease is partly due to selection effects 
(see Mannucci et al., in preparation, for a full explanation), 
but the observed trend
seems to be larger than what can be attributed to the effects of biases.
The presence of higher yields at lower stellar mass imply that
outflows decreasing with galaxy mass are not the main driver
of the mass-metallicity relation at z$>$3, and different possibilities,
related to the efficiency of star formation (e.g., \cite{brooks07}) 
or different outflow schemes (\cite{erb06}) must be considered.


%
%


\begin{thebibliography}{}

\bibitem[Brooks  et al. 2007]{brooks07}
{Brooks, A. M.,  et al.} 2007, \textit{ApJ}, 655, L17		

\bibitem[Chapman  et al. 2005]{chapman05}
{Chapman, S. C.,  et al.} 2005, \textit{ApJ}, 662, 772		

\bibitem[Conselice  et al. 2007]{conselice07}
{Conselice, S. C.,  et al.} 2007, \textit{MNRAS}, 386, 909	

\bibitem[Cowie  et al. 1996]{cowie96}
{Cowie, L.~L., Songaila, A., Hu, E.~M., \& Cohen, J.~G.}, 
1996, \textit{AJ}, 112, 839 					

\bibitem[de Rossi  et al. 2007]{derossi07}
{de Rossi, M. E., Tissera, P. B., \& Scannapieco, C.}, 
2007, \textit{MNRAS}, 374, 323					

\bibitem[Edmunds 1990]{edmunds90}
{Edmunds D.}, 1990, \textit{MNRAS}, 246, 678			

\bibitem[Erb  et al. 2006]{erb06}
{Erb D.}, 2006, \textit{ApJ}, 644, 813				

\bibitem[Franx  et al. 2003]{franx03}
{Franx, M.,  et al.}, 2003, \textit{ApJ}, 587, 79			

\bibitem[Garnett 2002]{garnett02}
{Garnett D.}, 2002, \textit{ApJ}, 581, 1019			

\bibitem[Grazian  et al. 2007]{grazian07}
{Grazian, A.,  et al.} 2007, \textit{A\&A}, 465, 393		

\bibitem[Kewley \& Ellison 2008]{kewley08}
{Kewley, L., \& Ellison, S. L.} 2008, \textit{ApJ}, 681, 1183	

\bibitem[Kobayashi  et al. 2007]{kobayashi07}
{Kobayashi, C., Springel, V., \& White, S. D. M.} 2007, 
\textit{MNRAS}, 376, 1465					

\bibitem[K\"oppen et al. 2007]{koppen07}
K\"oppen, J., Weidner, C., \& Kroupa, P. 2007, MNRAS, 375, 673	

\bibitem[Lee  et al., 2006]{lee06}
{Lee, H.,  et al.}, 2006, \textit{ApJ}, 647, 970		

\bibitem[Maiolino  et al. 2008]{maiolino08}
{Maiolino, R., et al} 2008, \textit{A\&A}, 329, 57 		

\bibitem[Mannucci  et al. 2002]{mannucci02}
{Mannucci, F.,  et al.} 2002, \textit{MNRAS}, 329, 57 		

\bibitem[Mannucci  et al. 2006]{mannucci06}
{Mannucci, F.,  et al.} 2006, \textit{MNRAS}, 360, 773		

\bibitem[Mannucci  et al. 2007]{mannucci07}
{Mannucci, F., et al} 2007, \textit{A\&A}, 461, 423		

\bibitem[Nagao  et al. 2006]{nagao06}
{Nagao, F.,  et al.} 2006, \textit{A\&A}, 459, 85			

\bibitem[Pozzetti  et al. 2007]{pozzetti07}
{Pozzetti, F.,  et al.} 2007, \textit{A\&A}, 474, 443 		

\bibitem[Savaglio  et al. 2005]{savaglio05}
{Savaglio, S.,  et al.} 2005, \textit{ApJ},  635, 260		

\bibitem[Steidel  et al. 2004]{steidel04}
{Steidel, C.,  et al.} 2004, \textit{ApJ}, 604, 534		

\bibitem[Tornatore et al. 2007]{tornatore07}
{Tornatore, L.  et al.}, 2007 \textit{MNRAS}, 382, 945		

\bibitem[Tremonti et al. 2004]{tremonti04}
{Tremonti, C. A.,  et al.}, 2004, \textit{ApJ}, 613, 898		

\end{thebibliography}
\end{document}